# Single-molecule imaging of lipid-anchored proteins reveals domains in the cytoplasmic leaflet of the cell membrane

Piet H.M. Lommerse[*†], Gerhard A. Blab[*], Laurent Cognet[*‡], Gregory S. Harms[*§],

B. Ewa Snaar-Jagalska[†], Herman P. Spaink[†] and Thomas Schmidt[*#]

[*]Department of Biophysics, Leiden University, Niels Bohrweg 2, 2333 CA Leiden, The Netherlands

[†]Department of Molecular Cell Biology, Institute of Biology, Leiden University, Wassenaarseweg 64, 2333 AL Leiden, The Netherlands

[‡] L. Cognet's present address is: Centre de Physique Moléculaire Optique et Hertzienne – CNRS UMR 5798 et Université Bordeaux 1, 351 Cours de la Libération, 33405 Talence, France

[§] G.S. Harms' present address is: Rudolf-Virchow-Center for Experimental Biomedicine, Julius-Maximilians University Würzburg, Versbacher Str. 9, 97078 Würzburg, Germany

**Corresponding author:**

[#] Dr. T. Schmidt, Department of Biophysics, Leiden University, Niels Bohrweg 2, 2333 CA Leiden, The Netherlands. Tel.: 31-71-527-5982, Fax.: 31-71-527-5819

E-mail: tschmidt@biophys.leidenuniv.nl.

**Manuscript information :** 27 text pages (including title page, abstract, text, acknowledgements, references, figure legends and table), 4 pages of figures, supporting information

**Word and character counts :** 178 words in the abstract, 46834 characters in paper

**Abbreviations footnote:** eYFP, enhanced yellow fluorescent protein; FRAP, fluorescence recovery after photobleaching

**ABSTRACT**

In the last decade evidence has accumulated that small domains of 30-700 nm in diameter are located in the exoplasmic leaflet of the plasma membrane. Most of these domains supposedly consist of specific sets of lipids- and proteins, and are believed to coordinate signal transduction cascades. Whether similar domains are also present in the cytoplasmic leaflet of the plasma membrane, is unclear so far. To investigate the presence of cytoplasmic leaflet domains, the H-Ras membrane-targeting sequence was fused to the C-terminus of the enhanced yellow fluorescent protein (eYFP). Using single-molecule fluorescence microscopy, trajectories of individual molecules diffusing in the inner leaflet of the plasma membrane were recorded. From these trajectories, the diffusion of individual membrane anchored eYFP molecules was studied in live cells on timescales from 5-200 milliseconds. The results show that the diffusion of 30-40% of the molecules is constrained in domains with a typical size of 200 nm. Neither breakdown of actin, nor cholesterol extraction changed the domain characteristics significantly, indicating that the observed domains may not be related to the membrane domains characterized so far.

## INTRODUCTION

The Singer & Nicholson fluid-mosaic model (1) has been the paradigm in cell biology for the description of the organization and dynamical behavior of the plasma membrane. In this model the membrane bilayer is represented as a neutral two-dimensional solvent in which membrane proteins are homogeneously distributed and are free to move. However in the last decade experimental evidence indicated that the plasma membrane of various mammalian cell types is heterogeneous in structure and contains various types of domains (2-5). One type of domain is referred to as liquid-ordered microdomain or 'lipid raft'. These cholesterol and sphingolipid-rich domains are thought to coexist with a more fluid phase and are enriched in phospholipids with saturated hydrocarbon chains (6, 7). Biochemically, lipid raft components are identified as the detergent-resistant membrane (DRM) fraction, observed after detergent extraction of cell membranes (8, 9). Various biophysical techniques were used to measure the size of lipid rafts *in vivo* yielding values between 30 and 700 nm (10-16). A second type of diffusional domain found in the membrane of live cells is caused by a cytoskeletal network associated with the cell membrane, resulting in domains with a diameter of 350-750 nm (3, 17).

Most of the biophysical studies to identify membrane domains have focussed on the exoplasmic leaflet of the plasma membrane, because of the easy accessibility of the outer membrane leaflet for the specific labeling of various membrane components. For this reason knowledge about domains in the cytoplasmic leaflet of the plasma membrane is immature, though recent work indicates their presence (16). From the biological point of view the cytoplasmic leaflet is of central importance, as most signaling pathways make use of proteins that are anchored in this leaflet, like small GTPases of the Ras family, the Scr-family kinases and G-proteins.

To investigate the presence of domains in the cytoplasmic leaflet a lipid-anchored fluorescent protein was produced which is targeted to the cytoplasmic leaflet of the plasma membrane. This protein consists of the enhanced yellow fluorescent protein (eYFP) which has the membrane targeting sequence of the human H-Ras fused to its C-terminus. This membrane targeting sequence consists of ten amino acids and contains three cysteines to which one S-prenyl and two S-palmitoyl groups are attached posttranslationally (18, 19). Association of such a lipid anchored fluorescent protein with lipid rafts has recently been suggested by biochemical methods where it has been found in the DRM fraction (20). However, other studies indicate that prenylated proteins are not localized in the DRM fraction (21, 22), making the possible association of prenylated proteins with lipid rafts unclear.

Because of the small size of the domains, advanced biophysical *in vivo* techniques are required for their identification and detailed study. Here we used single-molecule microscopy to track individual lipid-anchored fluorescent proteins in the cytoplasmic membrane of live cells. Detailed analysis of 35128 trajectories revealed that 30-40% fraction of the lipid-anchored eYFP is confined in domains. These domains are ~ 200 nm in size, are present in two different cell types, and at two different temperatures. The size of the observed domains does not depend significantly on the actin cytoskeleton or on cholesterol, which makes them different from membrane domains observed so far.

**MATERIALS AND METHODS**

**DNA-constructs**

The DNA sequence encoding the ten C-terminal amino-acids of human H-Ras (GCMSCKCVLS), which includes the CAAX motif, was inserted in frame at the C-terminus of the enhanced yellow-fluorescent protein (eYFP, S65G/S72A/T203Y) coding sequence using two complementary synthetic oligonucleotides (Isogen Bioscience, Maarssen, The Netherlands). The integrity of the reading frame of the eYFP-mtHR (mtHR for membrane targeting sequence H-Ras) was varified by sequence analysis. For expression in mammalian cells, the complete coding sequence of eYFP-mtHR was cloned into the pcDNA3.1 vector (Invitrogen, Groningen, The Netherlands).

**Cell cultures**

In this study two cell lines were used: a human embryonic kidney cell line (tsA201) and a mouse fibroblast cell line (3T3-A14). Cells were cultured in DMEM medium supplemented with streptomycin (100 µg/ml), penicillin (100 U/ml) and 10% bovine serum in a 7% $CO_2$ humidified atmosphere at 37°C (95% humidity). Cells were used for 12-14 passages and were transferred every 4 days. For microscopy the cells were cultured on #1 glass slides (Fisher Scientific, 's-Hertogenbosch, The Netherlands). Cells exhibiting a confluency level of 20% were used for transfection with either 2.5 µg DNA and 12.5 µl DOTAP (Amersham Pharmacia Biotech, Roosendaal, The Netherlands) or 1.0 µg DNA and 3 µl FuGENE 6 (Roche Molecular Biochemicals, Indianapolis, USA) per glass-slide. The transfection efficiency, as determined by fluorescence microscopy, was in the range of 10-30%.

**Single-molecule microscopy**

The experimental arrangement for single-molecule imaging has been described in detail previously (23, 24). Briefly, cells adherent to glass slides were mounted onto the microscope and kept in phosphate-buffered saline (PBS: 150 mM NaCl, 10 mM $Na_2HPO_4/NaH_2PO_4$, pH 7.4). The microscope (Axiovert 100TV; Zeiss, Oberkochen, Germany) was equipped with a 100× oil-immersion objective (NA=1.4, Zeiss, Oberkochen, Germany) and a temperature controller to keep the cells at 22 or 37°C. The samples were illuminated for 3 ms by an $Ar^+$-laser (Spectra Physics, Mountain View, CA, USA) at a wavelength of 514 nm. The illumination intensity was set to 2±0.2 $kW/cm^2$. Use of an appropriate filter combination (DCLP530, HQ570/80, Chroma Technology, Brattleboro, USA; and OG530-3, Schott, Mainz, Germany) permitted the detection of individual fluorophores by a liquid-nitrogen-cooled slow-scan CCD camera system (Princeton Instruments, Trenton, NY, USA). The total detection efficiency of the experimental setup was 8%. For the observation of the mobility of the membrane-anchored fluorophores the focus was set to the apical membrane of cells (depth of focus ~1 µm). The density of fluorescent proteins on the plasma membrane of transfected cells (4-6 days post transfection) was less than one per $\mu m^2$, which permitted imaging and tracking of individual fluorophores.

Fluorescence images were taken consecutively with up to 200 images per sequence. The signals on the CCD originating from individual molecules were fitted to a two-dimensional Gaussian surface with a full-width-at-half-maximum FWHM = 360±40 nm, given by the point-spread function of our setup. The photon counts were determined with a precision of 20%, limited by the shot-noise and readout-noise of the CCD camera. Comparison of the single-molecule signal with the background-noise yielded a signal-to-background noise ratio of 11. The latter figure translates into a positional accuracy for single-molecule localization of 35 nm (25). By connectivity analysis between consecutive images the two-dimensional trajectories of individual molecules in the plane of focus were reconstructed. These trajectories were up to 9 steps in length, limited by the blinking and photobleaching of the fluorophore (24). To compensate for the limited length of individual trajectories multiple data sets were produced. Each data set was acquired with a different time between two consecutive images (time lag, $t_{lag}$). By using different time lags, varying from 5 to 200 ms, the diffusion of individual molecules was studied.

**Trajectory analysis**

The trajectories were analyzed following a method developed earlier (26). In brief, the lateral diffusion of Brownian particles in a medium characterized by a diffusion constant D is described by the cumulative distribution function for the square displacements, $r^2$ (27, 28):

$$P(r^2, t_{lag}) = 1 - \exp\left(-\frac{r^2}{r_0^2(t_{lag})}\right) \qquad (1)$$

$P(r^2, t_{lag})$ describes the probability that the Brownian particle starting at the origin will be found within a circle of radius r at time $t_{lag}$. It is characterized by the mean-square displacement of $r_0^2(t_{lag}) = 4 D t_{lag}$ (27). Provided that the system under study segregates into two components, one with a fast and one with a slow mobility, characterized by diffusion constants $D_1$ and $D_2$, and relative fractions $\alpha$ and (1-$\alpha$), respectively, eq.(1) becomes (26):

$$P(r^2, t_{lag}) = 1 - \left[ \alpha \cdot \exp\left( -\frac{r^2}{r_1^2(t_{lag})} \right) + (1-\alpha) \cdot \exp\left( -\frac{r^2}{r_2^2(t_{lag})} \right) \right] \quad (2)$$

with mean-square displacements of $r_i^2(t_{lag}) = 4 D_i t_{lag}$.

The probability distributions $P(r^2, t_{lag})$ were constructed for every time lag from the single-molecule trajectories by counting the number of square displacements with values $\leq r^2$, and subsequent normalization by the total number of datapoints (26). Only probability distributions with N > 90 data points were least-square fit to eq.(2). This results in a parameter set { $r_1^2(t_{lag})$, $r_2^2(t_{lag})$, $\alpha$} for each time lag, $t_{lag}$, between 5 and 200 ms. By plotting $r_1^2$ and $r_2^2$ versus $t_{lag}$, the diffusional behaviour of the respective populations of molecules is revealed. The positional accuracy in our measurements is 35 nm, which leads to a constant offset in $r_i^2$ of $4 \times (35 \text{ nm})^2 = 0.49 \cdot 10^{-2}$ µm$^2$ for all time lags (26).

**Fluorescence recovery after photobleaching (FRAP)**

Cells were transfected with eYFP-mtHR 3-4 days before the experiment. During the experiments cells were kept in PBS at 22°C. FRAP curves (fig 5 supporting information) were fit to the equations given by Feder et al (29).

**Actin destabilization and cholesterol extraction**

The actin cytoskeleton was destabilized by supplementing the medium with 0.5 µM latrunculin B (Calbiochem, San Diego, CA, USA) followed by a 30 minute incubation period at 37°C and 7% $CO_2$. This treatment results in observable changes of the actin cytoskeleton (supporting information: fig. 6). As 0.025% DMSO is present during this latrunculin B treatment, control cells were incubated with 0.025% DMSO. After the incubation period, the cells were washed 3 times with PBS and measurements were taken within 30 minutes after incubation.

Cholesterol extraction was performed by incubating the cells in DMEM supplemented with 5 mM methyl-β-cyclodextrin (MβCD, Sigma-Aldrich Chemie, Steinheim, Germany) in a 7% $CO_2$ humidified atmosphere at 37°C (95% humidity) for 1 hour. This treatment decreases the free cholesterol content of the total cell membrane fraction by ~60 % (supporting information: fig. 7 and table 2). After the extraction cells were washed 3 times in PBS, and used for imaging for up to 1 hour.

**RESULTS**

**Observation and tracking of individual lipid-anchored molecules**

A fusion of the membrane targeting sequence of the human H-Ras with the enhanced yellow-fluorescent protein (eYFP-mtHR, see fig. 1A) was constructed in order to study its diffusion in the cytoplasmic leaflet of mammalian cells. Cells that were transiently transfected with DNA encoding this fusion protein showed a clear membrane-localized fluorescence two days after transfection (supporting information: fig. 8).

Four to six days after transfection, the density of membrane anchored eYFP molecules was low enough ($< 1$ $\mu m^{-2}$) to observe signals of individual eYFP-mtHR molecules (fig.1A). Diffraction limited spots with a mean signal intensity of 73 ± 14 cnts/ms, and single-step photobleaching events (fig.1B-C) ensured that individual fluorescent molecules were observed (23, 24). Consecutive illuminations of the flat apical membrane area were used for tracking of individual eYFP molecules and construction of corresponding trajectories. In fig.2A two trajectories taken with a time lag of 80 ms are shown. The positional accuracy with which single molecules could be localized was 35 nm. Variation of the time between two images allowed recording of sets of trajectories with time lags ranging from 5 to 200 ms.

For each time lag, a set of trajectories (N>90) was used to calculate the cumulative probability distribution of the square displacement. Figure 2B shows the cumulative probability distribution for the time lag of 64 ms (dots). The data exhibit a biphasic behavior described by the bi-exponential probability function given in eq.(2) (solid line). For the data shown in fig.2B the diffusion of molecules was characterized by a relative fraction, α = 0.76±0.05 and the mean square displacements $r_1^2 = 0.16±0.02 \, \mu m^2$ and $r_2^2 = 0.012±0.005 \, \mu m^2$, respectively. For all time lags, samples, and environments measured, the data exhibit such biphasic behavior.

**Single-molecule measurements on eYFP-mtHR in tsA201 cells**

Single-molecule diffusion measurements were first performed on eYFP-mtHR in tsA201 cells at 22°C. Data sets with time lags between 8 and 200 ms were obtained and analyzed as described in the previous subsection, yielding the corresponding mean square displacements and fractions. The data are summarized in fig. 3A-C and table 1. The fast diffusing molecules, characterized by the mean square displacement $r_1^2$, represent the largest fraction of molecules (73±5%). The fraction stays constant for time lags between 8 and 200 ms (fig.3A). The mean square displacement of this fast fraction followed a linear increase with time predicted for a freely diffusing species, $r_1^2 = 4 D_1 t$ (figure 3B, solid line) and is characterized by a diffusion constant of $D_1 = 0.53±0.10 \, \mu m^2/s$. The other fraction, containing 27±5% of the molecules, exhibited a reduced mobility (fig. 3C). Below 75 ms the mean square displacement increased with time, however, leveled off to a constant value of $1.9±0.6 \cdot 10^{-2} \, \mu m^2$ for longer time lags.

Such asymptotic behavior is explained by a confined diffusion model. The model assumes that diffusion is free within a square of side length L, which is surrounded by an impermeable, reflecting barrier. In such a model the mean square displacement depends on L and the initial diffusion constant $D_0$, and varies with $t_{lag}$ as (30):

$$r_2^2(t_{lag}) = \frac{L^2}{3} \cdot \left(1 - \exp\left(\frac{-12 D_0 t_{lag}}{L^2}\right)\right) \qquad (3)$$

From a fit of the data to eq.(3) (solid line in fig.3C) we obtained an instantaneous diffusion constant of $D_0 = 0.08\pm0.02$ µm$^2$/s and an average domain size of $L = 241\pm35$ nm.

**FRAP on eYFP-mtHR in tsA201 cells**

In addition to single-molecule microscopy we performed fluorescence recovery after photobleaching (FRAP) experiments to determine if the domains seen in the single-molecule experiments would show up as an immobile fraction in FRAP. Recovery curves (N=8, supporting information fig. 5) were fit to the model described in Feder et al. (29). The analysis yielded an average diffusion coefficient $D_{FRAP} = 0.47\pm0.17$ µm$^2$/s and a mobile fraction of 74±12 %. The immobile population of 26±12 % in the FRAP experiments indicates that the domains observed in the single-molecule experiments are stable on a timescale of tens of seconds (for comparison see in table 1).

**Single-molecule measurements on eYFP-mtHR in 3T3-A14 cells**

To investigate if the domains observed in tsA201 cells were also present in a different cell line, we conducted single-molecule diffusion measurements on a mouse fibroblast cell line (3T3-A14) at 37°C. Data sets with time lags between 5 and 60 ms were obtained. The data are summarized in fig. 3D-F. The fast diffusing fraction again contains most molecules, 59±7%, and stays constant between 10-60 ms (fig. 3D, table 1). The corresponding mean square displacement followed a linear increase with time (fig. 3E, solid line), characterized by a diffusion constant $D_1 = 1.13\pm0.09$ µm$^2$/s. The minor fraction (41±7%) of molecules exhibited a confined diffusion behavior. Fit of

eq.(3) to the data yielded an instantaneous diffusion coefficient $D_0$ of $0.29\pm0.12$ µm$^2$/s and an average domain size L = $206\pm35$ nm.

**Involvement of actin and cholesterol in the observed domains**

To elucidate the nature of the observed domains, two possibilities were investigated: the involvement of the actin cytoskeleton and the hypothesis that the domains were associated with cholesterol dependent liquid-ordered lipid domains. Actin dependence was tested by treating 3T3-A14 cells with 0.5 µM of latrunculin B, which disrupts microfilament organization by the formation of a 1:1 complex with monomeric G-actin (see supporting information fig.6 ). Single-molecule measurements (37$^o$C) and diffusion analysis again revealed a fast, free diffusing population and a slower confined diffusing population of molecules (fig. 4A-C, table 1). The diffusion coefficient of the slow fraction ($D_0 = 0.31\pm0.17$ µm$^2$/s), as well as the domain size (L = $177\pm35$ nm), were not significantly different from the untreated (fig. 3D-F) or DMSO treated controls (dashed lines in fig 4A-C). However, it should be noted that the addition of 0.025% DMSO significantly reduced the diffusion constant of the free diffusing fraction from $1.13\pm0.09$ µm$^2$/s to $0.83\pm0.05$ µm$^2$/s.

The possibility that the observed domains were cholesterol dependent liquid-ordered domains, was tested by incubation of 3T3-A14 cells with 5 mM methyl-β-cyclodextrin for 1 hour at 37$^o$C. This treatment results in the extraction of 60% of the cholesterol out of the cellular membranes (supporting information: fig. 7 and table 2). Single-molecule imaging and diffusion analysis was performed (fig. 4D-F), which again revealed two fractions of diffusing molecules. The diffusional behavior of these two fractions ($D_1 = 0.95\pm0.06$ µm$^2$/s, $D_0 = 0.14\pm0.08$ µm$^2$/s) and domain size (L = 204±67 nm) did not differ significantly from untreated cells (dashed lines in fig 4D-F; table 1). However, the freely mobile fraction increased to $\alpha = 72\pm10$ % (control: 59±7%) after cholesterol extraction.

**DISCUSSION**

Our data show that diffusional domains in the cytoplasmic leaflet of the plasma membrane in mammalian cells exist and that proteins anchored to the plasma membrane via the H-Ras membrane anchoring sequence are only partially captured within those domains. The average size of the observed domains was 241±35 nm for human-embryo kidney tsA210 cells. It is interesting to note that the fraction of molecules which exhibits confined diffusion in the single-molecule experiments (27 %) in these cells, corresponds well to the immobile fraction observed in our FRAP experiments (26 %). This similarity strongly suggests that the domains, which are much smaller than the bleaching spot in FRAP (200 nm vs 1800 nm), are stable on the timescale of the FRAP recovery lasting up to 10 seconds. Similar, though smaller (10-13%), immobile fractions have been observed in FRAP experiments on Rat-1 cells expressing a GFP-H-Ras fusion protein (31).

Domains were observed for both mouse fibroblast 3T3-A14 cells and human-embryo kidney tsA201 cells. The size of the domains was independent of cell type. The fraction of molecules inside these domains was significantly larger in 3T3-A14 cells compared to tsA201 cells (41±7% vs. 27±5%), which could be due to cell type related differences and the different temperatures used. As expected, the diffusion coefficients measured at 37$^{o}$C are higher compared to those measured at 22$^{o}$C.

In order to further elucidate the nature of the observed domains, 3T3-A14 cells were treated with drugs that would potentially disrupt domain organization. Under the assumption that the observed domains are liquid-ordered domains, extraction of 60% cholesterol from the cell membrane should result in a decrease in the size of the observed domains, a decrease of the population of molecules in domains, or a combination of these two effects. This has not been found in our experiments. The fact that the eYFP-mtHR molecules do not partition in liquid-ordered domains to a significant extent is not completely unexpected. The presence of a branched and multiply unsaturated farnesyl group on the membrane targeting sequence of H-Ras (figure 1) does not favor a high partitioning in liquid-ordered domains (21, 22, 32, 33).

This was confirmed by a recent study using fluorescence resonance energy transfer (FRET), which revealed that geranylgeranylation only, does not promote clustering in cholestrol- and sphingolipid-rich domains (16), but results in cholesterol independent clustering. Palmitoylation is reversible and dynamic (34) (35), so it cannot be ruled out that a fraction of the observed eYFP-mtHR molecules is not fully palmitoylated, resulting in a localization in non-cholesterol dependent clusters as observed by Zacharias et al (16).

However in recent biochemical studies, where the complete H-Ras membrane-targeting sequence (as used in the current study) was fused to GFP (GFP-mtHR) it was found in the DRM fraction (20), indicating the potential affinity for liquid-ordered domains. Additionally, a recent electron microscopy (EM) study showed that GFP-mtHR is localized in cholesterol dependent domains with a mean diameter of 44 nm that occupy as much as 35% of the cells surface (36). A similar study showed that 44% of the GFP-mtHR is localized in caveolae (20). A direct comparison of the EM-data with the current diffusion study is difficult. As the EM results were obtained on fixed plasma membrane fragments the pre-fixation structure and dynamical behavior is problematical to infer. An association of the domains found in the present study with those detected by EM seems an attractive possibility. The lack of any cholesterol-dependence in the present study makes this link less likely. However, we cannot entirely exclude that the structures observed in EM do partially account for the domains found here.

In order to elucidate the possible role of the cortical actin on the organization of the observed domains we applied the actin-depolimerization drug latrunculin B. However, the cytoplasmic leaflet-domains observed in this study were not sensitive to latrunculin B, making it unlikely that they are due to the membrane-skeleton fence as proposed by Kusumi and Sako (30) (17).

As to the biological function of the observed domains, it was speculated that domains might play a role in separating H-Ras molecules in the inactive (GDP-bound) state, from H-Ras molecules in the active (GTP-bound) state (20). Recent biochemical data (20), indicate that cholesterol dependent lipid rafts play a major part in this separation function. The results and techniques described in this paper provide a starting point to directly investigate the involvement of membrane domains in signaling processes *in vivo* with high spatial and temporal resolution.

**ACKNOWLEDGEMENTS**

We thank J.Y.P. Butter for help with FRAP experiments and G.E.M. Lamers for assistance with the control experiments and confocal microscopy. Furthermore, we like to thank A.A. de Boer for maintenance of the cell cultures. The 3T3-A14 cells were a generous gift from Dr. J.A. Maassen, Leiden University Medical Center. This work was supported by funds from the Dutch ALW/FOM/NWO program for Physical Biology (99FBK03). L.C. acknowledges support from DGA/DSP (France) and the European Marie-Curie fellowship program (IHP-MCFI-1999-00736).

**FIGURE 1 (A)** Schematic drawing of the eYFP-mtHR protein including the S-prenyl and two S-palmitoyl groups. White light (upper left) and corresponding fluorescence images (upper and lower right) of a tsA201 cell transfected with DNA encoding for eYFP-mtHR. For the fluorescence image the cell was illuminated using 514 nm light for 3 ms at an intensity of 2 kW/cm$^2$. Two membrane-localized signals attributed to single fluorescent proteins are present. **(B)** Example of a single-step photobleaching event of an individual eYFP-mtHR, indicative for an individual fluorophore. **(C)** Analysis of 240 signals of individual eYFP-mtHR molecules observed at the apical membrane of tsA201 cells (solid line). The probability density of the signal amplitude is nearly Gaussian-shaped with a maximum of 220 cnts/3ms. The statistics of the background signal (dashed line) is shown for comparison, being characterized by a width of $\sigma_B$ = 19 cnts/3ms.

**FIGURE 2 (A)** Example of two trajectories of individual eYFP-mtHR molecules at the apical membrane of a tsA201 cell. The time between subsequent observations was 80 ms. **(B)** Cumulative distribution function for square displacements (N = 290), $P(r^2, t_{lag})$, of individual eYFP-mtHR molecules observed at the apical membrane of ten different tsA201 cells. Data were obtained with a time lag of 64 ms. The solid line represents the result of a bi-exponential fit according to eq.(2), yielding $r_1^2$ = 0.16±0.02 μm$^2$, $r_2^2$ = 0.012±0.005 μm$^2$ and α = 0.76±0.05. A mono-exponential fit according to eq.(1) (dashed line) fails.

**FIGURE 3** Results obtained from the square displacement distribution analysis (according to fig. 2B) of data taken on tsA201 cells at 22°C (A-C), and data taken on 3T3-A14 cells at 37°C (D-F). The results of the fits are given in table 1. The error bars represent the standard errors obtained from the fits of the data according to eq. 2. **(A and D)** Fractions of the fast component, α, versus $t_{lag}$ for the two cell types studied. **(B and E)** The mean square displacements, $r_1^2$ of the fast fraction versus $t_{lag}$. The data are fitted according to a free diffusion model ($r_1^2 = 4D_1 t_{lag}$, solid line). **(C and F)** Mean square displacements, $r_2^2$ of the slow fraction versus $t_{lag}$. The data are fitted according to a confined diffusion model (eq.(3), solid line). The dotted lines in fig.3C and F represent the offset due to the limited positional accuracy (see materials and methods).

**FIGURE 4** Results obtained from the square displacement distribution analysis of data taken on 3T3-A14 cells at 37°C after drug treatment. The fit-results obtained for DMSO treated cells (fig. 4A-C) and untreated cells (fig. 4D-E) are shown for comparison as dashed lines, results from the fits are given in table 1. **(A-C)** Fraction (fig.4A), and mean-squared displacement of the fast (fig. 4B) and the slow (fig. 4C) fraction after treatment with 0.5 µM latrunculin B for 30 min. **(D-F)** Fraction (fig.4D), and mean-squared displacement of the fast (fig. 4E) and the slow (fig. 4F) fraction after treatment with 5 mM methyl-β-cyclodextrin for 1 hour.

**TABLE 1 Summary of diffusion characteristics**

|  | tsA201 cells, 22 °C | | 3T3-A14 cells, 37 °C | | | |
|---|---|---|---|---|---|---|
|  | single-molecule control | FRAP control | single-molecule | | | |
|  |  |  | control | + DMSO | + lat B | + MβCD |
| $\alpha$ | 0.73±0.05 | 0.74±0.12 | 0.59±0.07 | 0.71±0.10 | 0.67±0.05 | 0.72±0.10 |
| $D_1$ ($\mu m^2/s$) | 0.53±0.10 | 0.47±0.17 | 1.13±0.09 | 0.89±0.04 | 0.83±0.05 | 0.95±0.06 |
| $D_0$ ($\mu m^2/s$) | 0.08±0.02 |  | 0.29±0.12 | 0.23±0.12 | 0.31±0.17 | 0.14±0.08 |
| L (nm) | 241±35 |  | 206±35 | 213±45 | 177±35 | 240±67 |

**Fig.1**

**A**

10μm

0  cnts/pxl  100

**B**

signal (counts/3ms)

time (ms)

**C**

ρ (counts/3ms)$^{-1}$

(x 5)

x 10$^{-3}$

signal (counts/3ms)

**Fig.2**

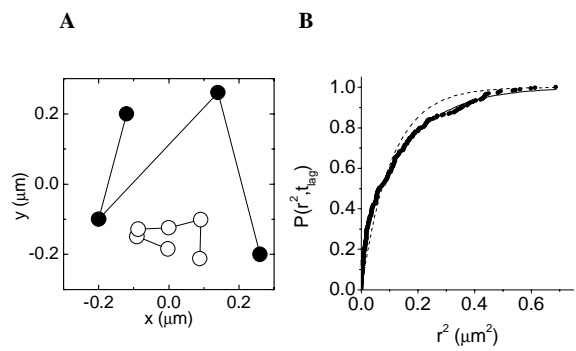

**Fig.3**

**tSA201 cells at 22°C**

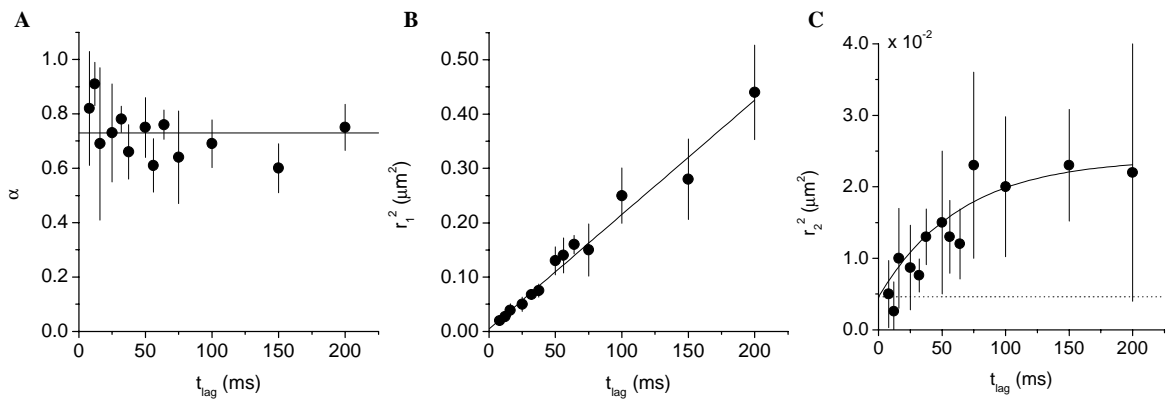

**3T3-A14 cells at 37°C**

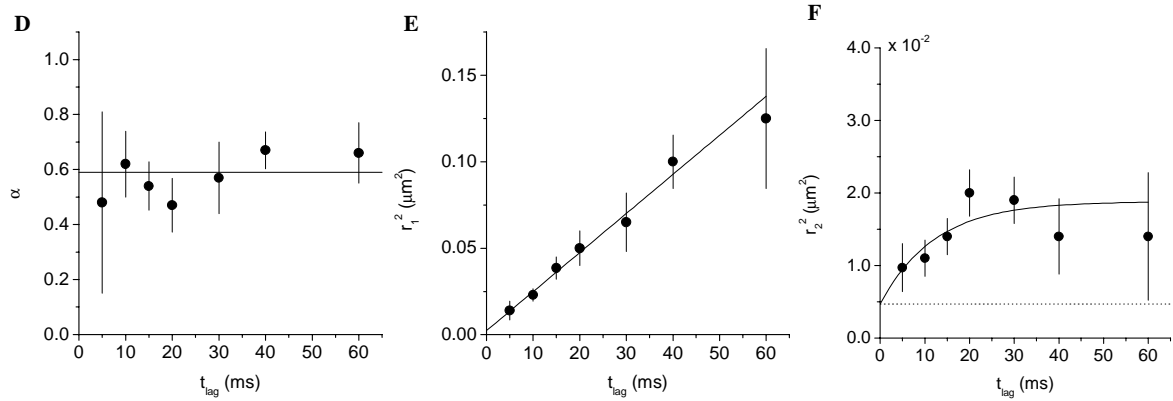

**Fig.4**

**Latrunculin B**

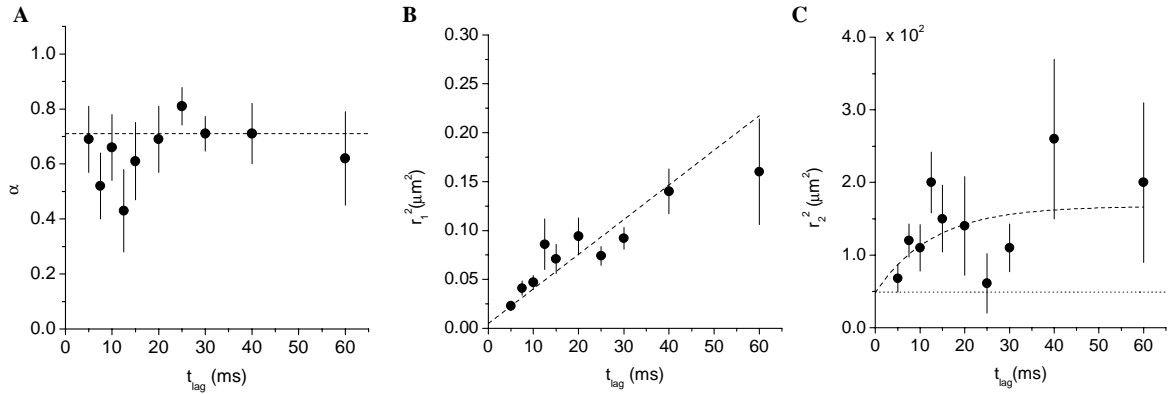

**MβCD**

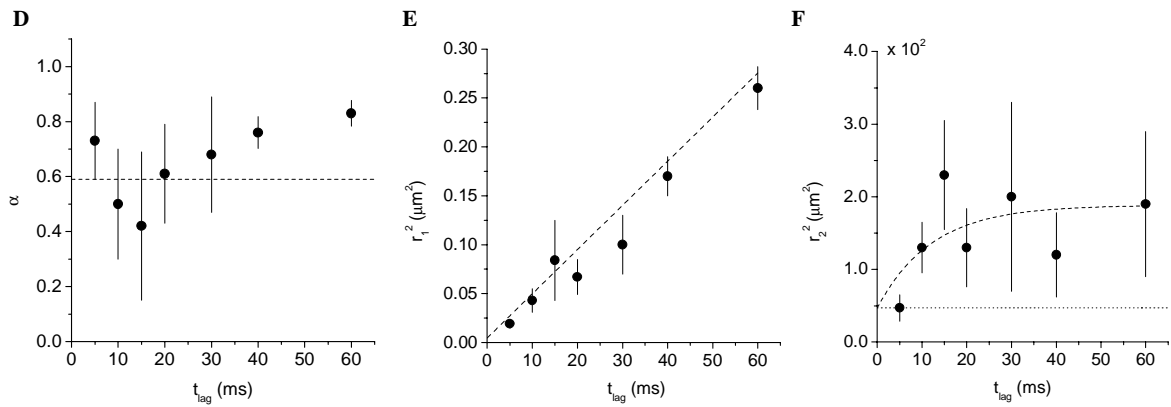